\documentclass[11pt,twoside]{article}
\usepackage{asp2004}
\usepackage{psfig}
\usepackage{epsf}
\usepackage{graphics}
\usepackage{lscape}
\def\lesssim{\mathrel{\hbox{\rlap{\hbox{\lower4pt\hbox{$\sim$}}}\hbox{$<$}}}}
\def\gtrsim{\mathrel{\hbox{\rlap{\hbox{\lower4pt\hbox{$\sim$}}}\hbox{$>$}}}}

\def\omegam{{$\Omega_{\rm m}$}}

\def\omegab{{$\Omega_{\rm b}$}}

\def\omegal{{$\Omega_\Lambda$}}
 
\def\omegabh2{{\omegab h^2}}

\begin{document}

\title{Nearby Large-Scale structures and the Zone of Avoidance: A Conference Summary and Assessment}

\author{A.P. Fairall} 
\affil{Department of Astronomy, University of Cape Town, Private Bag,
        Rondebosch 7700, South Africa}

\author{Ofer Lahav}
\affil{Department of Physics and Astronomy , University College London, 
Gower Street, London WC1E 6BT, UK\\}

\begin{abstract}
This conference has brought together many of the world's
`cosmographers', particularly those focussed on the mapping nearby
large-scale structures. Not surprisingly, numerous portrayals of the
Local Cosmic Web and its characteristics have been presented, with
some reconciliation to the structures found by the much deeper 2dF
Galaxy Redshift Survey and the Sloan Digital Sky Survey. In recent
years, Near-infrared and H I surveys have greatly narrowed the `Zone
of Avoidance' caused by the foreground obscuration of the Milky
Way. There has been emphasis towards southern large-scale structures,
with the particular need to identify the mass overdensities
responsible for the dipole in the Cosmic Microwave Background. While
the general consensus is that the Shapley Concentration is a greater
overdensity than the Great Attractor, its far greater distance makes
it unclear as to whether its gravitational effect on the Local Group could be
larger. Furthermore, worries about residual bulk motion remain.
\end{abstract}

\section{Introduction - the big picture}

This written report of the conference summary is based on the framework presented orally by one of us (Lahav) at the conclusion of the proceedings.

The conference has concentrated on the nearby Universe, particularly that
mapped and explored by redshift surveys. It must, however, be seen as 
part of a big picture, a quest for knowledge of the Universe in general. Obviously whatever emerges on a local scale must be compatible with evolution that has taken place since the embryonic Universe recorded its imprint upon the Cosmic Microwave Background (CMB). The Universe on very large 
scales is also being explored by means of probes such as Type Ia supernovae and lensing, which give considerable insight into the nature and behaviour of the Universe in the past. What we see as nearby large-scale structures, and the galaxies they contain, is the universe today, the outcome 
of its past evolution.

At this conference we have seen a distinction between `cosmographers'
and `statisticians'. Cosmographers are focussed - some might say
obsessed - on the mapping of individual local large-scale
structures. Admittedly, such mappings are vital to the understanding
of local peculiar motions (about which more will be said
below). Statisticians on the other hand are more concerned with the
general characteristics of large-scale structures, particularly the
Power Spectrum of density fluctuations, which is seen as the prime tool for interface with
that from the CMB. Putting power spectra from both local surveys and
the CMB together gives us very strong constraints on various
cosmological parameters. The most obvious recent example is the 2dF
survey. As reported by one of us (Lahav) at the
conference~\footnote{All references in this article are to presenters
at the conference, who appear as first authors elsewhere in these
proceedings.}, 2dF has given us \omegam = 0.3, an upper limit on the
neutrino mass, and more.  Early results from the Sloan Digital Sky
Survey (SDSS) are in concordance with the 2dF results.  Both surveys
support the `concordance model' of a flat universe with approximately
\omegab = 0.04, \omegam = 0.26, \omegal = 0.70 and a Hubble parameter
of approximately 70 km s$^{-1}$ Mpc$^{-1}$.

The old question of light versus mass has surfaced repeatedly, particularly when mass overdensities are associated with rich clusters. Do the optically visible galaxies, that are the data points in the redshift surveys and which define the large-scale structures, also show where mass is concentrated? Evidence such as the determination of \omegam$^{0.6}/b$ = 0.55 $\pm$ 0.06 (Lucey), give a bias parameter b near unity, thereby supporting the general belief that luminous and dark matter go together, though clearly the dark matter is nowhere as condensed as the visible baryonic matter. But we have some confidence from the studies of peculiar velocities where galaxies are downgraded to test particles moving under the influence of the mass concentrations. Those studies reveal mass concentrations that generally coincide with the visible large-scale structures.

The long ongoing controversy over the value of the Hubble parameter has happily moved towards consensus with $H_{0} = 72 \pm 8$ km s$^{-1}$ Mpc$^{-1}$ (although conference participants still made use of $h_{50}, h_{70}, h_{75}$ and $h_{100}$). Feast reviewed the status of the primary distance indicators, upon which our knowledge of the scale of the Universe rests. As he points out, one group using HST data is still able to get H$_{0}$ = 60 km s$^{-1}$ Mpc$^{-1}$ and there is little chance of being able to resolve this further until there are more and better observations from space. 

\section{The Zone of Avoidance}

Appropriately, the opening function of the conference took place at the site where Sir John Herschel set up his famous telescope to chart the southern skies. Not only did Herschel recognise the first large-scale structure (the Virgo Supercluster), but he also called attention to the Zone of Avoidance (ZoA). While the ZoA is inconvenient to seeing the position of our Galaxy in terms of local cosmography, it has nevertheless given employment to most of the conference participants, and a reason to hold these conferences!

It is ten years since the first of these meetings was held, and it is pleasing to report on remarkable progress. Near-infrared surveys (as reviewed here by Kraan-Korteweg) are far less affected by the extinction of the Milky Way's  dust; while optical surveys are ineffective beyond $A_{B} = 3^{m}$, near-infra surveys penetrate to $A_{B} \sim 10^{m}$. Progress has come from both the PSC-z Behind the Plane Survey (Saunders) and the highly successful 2MASS survey (as reported by Huchra). Similarly, blind H I surveys, which are totally unaffected by the foreground Milky Way, have advanced enormously, thanks to the Multi-Beam receiver on the Parkes Telescope (see Henning, Staveley-Smith). For completeness, a similar survey must still be carried out in the northern skies.

These combined efforts have greatly narrowed the ZoA, such that we
have confidence that we are no longer missing significant nearby
large-scale structures. For a start, there is no M31-like galaxy
hidden behind the Milky Way. A number of new galaxies and clusters
(e.g. those mentioned below) have been detected in the ZoA. The H I
studies in particular have revealed the connectivity of large-scale
structures across the ZoA (see Koribalski). As yet this work has
provided mappings; there is still a need to quantify what has been
discovered, so that structures in the ZoA can be more directly
compared to those at higher latitudes.

Optical surveys and photometry nevertheless still serve a complimentary function. This is particularly true where the foreground star density of the Galactic bulge is high enough to confuse surveys such as 2MASS (see Kraan-Korteweg: Optical to NIR studies). Wakamatsu reports on his mapping of the Ophiuchus Cluster, seen through the bulge, and its extension into an apparently massive and influential supercluster at cz $\sim$ 9000 km s$^{-1}$.

In optical surveys, knowledge of reddening is essential. There has also been good progress in this regard (see Burstein: Reddening in the ZOA); though, where extinction is high, the resolution of the Burstein \& Heiles and the Schlegel et al. maps is still not quite adequate for individual galaxies,. The problem has also been inverted by Gonzalez et al., who derive extinctions through disk galaxies, and particularly address the crowded star fields of the Galactic bulge. As long as gravitationally-influential structures seem to lie within the ZoA, there is plenty of work cut out to do.

\section{The Local Cosmic Web}

Nearby large-scale structures - forming the Local Cosmic Web - were also the focus of the meeting.

Starting closest to home, Karachentsev et al. presented their new `Catalogue of Nearby Galaxies', the closest approximation we currently have to a volume-limited sample. One of their plots reveals the most local of filaments forming a network over half the sky, the other half dominated by the Local Void and a `Mini Void' (the Cetus Void). Some of these nearby galaxies are in for detailed scrutiny via the VLA/Spitzer SINGS survey (de Blok et al.) which will investigate key characteristics of galaxy morphology. It is also these closest galaxies that indicated to Fairall et al. that every galaxy, within the same large-scale structure, has a neighbour within 100 km/s of redshift space. All galaxies lie in filaments or tree structures.

The mapping of filaments is very obvious in the H I observations of the HIPASS Survey, as seen in the deeper maps of Koribalski. This survey has picked up the population of (optically) fainter galaxies that escapes inclusion in magnitude-limited surveys, particularly higher redshift surveys such as 2dF and Sloan. When such galaxies are included, filaments appear sharper and connectivity between large-scale features is improved. A web of filaments seems to be the accepted character of the local Universe. Nevertheless one of us (Fairall) has pointed out that classic `bubbly' texture still exists, but that filaments of galaxies surround the bubbly voids, much like the filaments in the shells of supernova remnants.

A presenter who, in passing, casually referred to `field galaxies' was immediately shouted down. There was subsequently clear agreement amongst the conference participants that there is no such thing as field galaxies. The hierarchy of clustering, and the interconnecting of filaments, ensures that every galaxy belongs to some or other large-scale structure.

Within our Local (Virgo) Supercluster, the Supergalactic Plane is still clearly recognised and used as reference (It is a pity that Gerard de Vaucouleurs, who first defined it, is not still around to witness his success in this regard). Beyond the Local Supercluster is the Great Attractor region. It is interesting to note that some (Tully, Fairall) saw our Local Supercluster as an appendage of the Great Attractor/Centaurus region, while one very significant person (Huchra) did not. Clearly there are still details of the local cosmography that need to be sorted out.

Voids also received attention, particularly the Local Void, which was the focus of the investigation by Iwata, who claimed very significant peculiar velocities for galaxies on the far side, indicating the expansion of the void. The Local Void also received particular mention by Karaschentsev, Koribalski, and Tully, and came up for frequent discussion. There are differences in how voids are perceived. Some, working on very nearby structures (esp. Fairall) see them as empty, other working on the more distant (lighter sampled) data see them as underdense regions (e.g. Einasto).

A stunning sky-wide deeper view of the local universe was provided by the 2MASS survey (Huchra). Currently almost 23 000 redshifts are available for 2MASS galaxies, thereby allowing a variety of plots (hockey pucks and onion skins) showing data to cz $\sim$ 10\,000 km s$^{-1}$. The database has been also been expanded enormously by a technique that infers redshifts (developed by Jarrett, who regrettably was unable to be at this meeting). 2MASS also supplies the targets for the 6dF survey (Jones) which is prolifically producing redshifts, at least in the southern skies. This survey aims to get redshifts of 170 000 galaxies, and peculiar velocities of a 15 000 subset, by far the largest redshift survey of the local universe.

2MASS is probably as deep as we can see large-scale structures over the whole sky, but 2dF and SDSS are carrying the mappings far deeper. Lahav reported how the underlying density field of 2dF has been reconstructed by means of Wiener filtering. Similarly Einasto et al. have looked at the structures in SDSS (Data Release 1) and identified both clusters and superclusters; they find clusters in a high-density environment to have about five times the luminosity of clusters in low-density environments. Although most SSDS galaxies are distant and HIPASS/HIJASS galaxies nearer, Garcia-Appadoo presented a preliminary analysis of galaxies common to both surveys, offering insight into star formation processes.

Unfortunately, these deeper surveys do not include the lower Galactic latitudes of the Shapley Supercluster, which (as discussed below) now appears to be the most influential of local overdensities. Clearly deeper mappings are badly needed in this direction.

It could also be that we are missing very large-scale local inhomogeneities. Frith reported that the APM survey area may be 25 percent deficient in galaxies out to z $\sim$ 0.1. This is partly based on models guided by the 2dF survey. The other long established probe of very-large-scale inhomogeneities are the Abell Clusters, which received attention from Andernach et al. Such clusters have long since been considered as beacons of large-scale structures. From the smaller-scale end, Andernach et al. succeed in marrying compact groups (like the well-known Hickson groups) to Abell clusters.

New tools are needed in examining local large-scale structures. The classic N-body simulations, that serve so well representing the larger-scale features and the derived power spectrum, need to progress down to smaller scales. Clearly we need simulations that create both the large-scale features, and individual galaxies as well. Weak lensing has clearly emerged as a powerful tool (although it did not feature in any of the conference presentations) that will track the irregularities in the mass distribution.

Since we cannot see the same structures at different eras, quantitative statistic measures are still the key to reconciling structures seen today to those in the past. In this regard, the favoured measure is still the Power Spectrum (or sometimes its Fourier transform, the Correlation Function). Alternative statistical measures are desirable, particularly those sensitive to texture (the Correlation Function `industry' was unaffected years ago by the discovery of voids, since it is blind to texture). One of us (Lahav) has also made the point that two realisations of the galaxy distribution may have the same power spectrum, yet look very different, due to phase correlations. Such phase information is generally discarded when power spectra are derived.

Indeed the 'texture' of the universe is sometimes best brought home by visualisations, which are also essential for showing up local cosmography. Examples were seen at this conference (Tully, Fairall). What is still needed is the ability to display N-body simulations in the same fashion, side-by side with the real data.

\section{The Dipole and bulk motions}

One cannot look at large-scale structures without the involvement of large-scale motions, in particular the cause of the motion of the Local Group of galaxies with respect to the Cosmic Microwave Background. Studies of the ZoA have been hastened by the proximity of the Dipole to the Galactic Plane. Is the Great Attractor at rest, or does it too have a peculiar motion in the same direction as ours? Woudt reviewed this problem, and offered hope that fundamental plane analysis of the Norma Cluster may provide the answer.

What was obvious at the conference was a tendency to downplay the role
of the Great Attractor, as compared to more distant structures. While
there was agreement that the more distant Shapley Concentration was a
more significant mass overdensity, its mass would have to be more than
an order of magnitude greater than that of the Great Attractor for it
to have the same effect at the Local Group; such is not apparent in optical, IRAS and
2MASS redshift surveys. Nevertheless, Lucey used supernovae to find
strong support for a sizable flow towards the Shapley Concentration.

The X-ray community at the conference (Ebeling, Mullis and Kocevski) pushed hard on the fact that significant potential wells can be traced by the hot gas responsible for diffuse X-ray sources. Almost all such overdensities out to z $\sim$ 0.07 can be found in this way. (On a more local scale, Burstein also reported hot gas even in groups of galaxies.) The X-ray evidence lies strongly in favour of the more distant Shapely Concentration and similarly distant clusters in Triangulum Australis (on the other side of the Galactic Plane) being mainly responsible for the Dipole.

The Great Attractor Region was not however ignored. No less than three papers (Kraan-Korteweg, Schroeder, Nagayama) were devoted to the Cluster around PKS 1343-601 (with redshift matching the GA). All three investigations found a significant cluster almost totally obscured by the Milky Way, though nowhere as massive as the Norma Cluster. Surprisingly the X-ray observers (Ebeling and Mullis) did not support their finding, as they reported only anaemic X-ray emission from that cluster, and rather favoured the nearby cluster CIZA J1324 as a more prominent overdensity. While the optical and X-ray researchers were not really in agreement on the relative roles of clusters in the Great Attractor region, there was nevertheless consensus that the Great Attractor is no longer the major role player it was once thought to be.

Einasto et al. also reported on N-body simulations in this regard: Very massive superclusters form `Great Attractors' and have small bulk motions. Less massive superclusters form smaller attractors, and have larger bulk motions.

More worrying is the nagging problem as to whether bulk motions occur on even larger scales. Has the Postman-Lauer effect gone away? Michael Hudson reviewed the search for a volume of space which is at rest. He discussed and explained the differences between the various surveys of the peculiar motions of clusters, almost all showing disturbing large bulk motions in the general direction of the Shapley region, though apparently towards a more distant target. Combining all these surveys results in a flow of 350 km/s. Does this indicate the existence of a `very large-scale structure' still more distant. In ten years, the problem has not gone away.

The hope then lies with deeper probes and newer, better tools. The 6dF survey is going to make a major contribution to knowledge of local flows (Jones) but may not go deep enough. Unfortunately it does not give coverage of the northern hemisphere.

New methods of distance determination, capable of operating at deeper distances, are obviously needed: Dn-Sigma, Surface brightness fluctuations, and the kinematic Sunyaev-Zeldovich effect. On the theoretical side, there is much need to model the non-linear velocity field with the aid of N-body simulations.

\section{The Future}

The almost exponential increase in the production of redshifts allows us to make enormous improvements in the statistics. This applies not only to the large-scale, but also to the most local scale where we can begin to approximate volume-limited samples that will give us the detailed characteristics of small scale formation.

What we have found with nearby large-scale structures must still be reconciled with the `big picture'. The 2dF and SDSS have clearly revealed the bimodality of galaxy populations. The clustering patterns of these two different populations are different, and are therefore telling us something fundamental about how galaxies and large-scale structures have formed and evolved.

For the first time, we have the ability to discern clustering patterns at z $\sim$ 1, with the DEEP2 and VIRMOS surveys. The advent of the Southern African Large Telescope (Buckley) offers an unprecedented opportunity to local observers.

In parallel to this exploration, we need to match our observations to N-body simulations. In so doing, we may trace the fundamental physics that shapes the fabric of the Universe.

\end{document}